# *Ion Transit Pathways and Gating in ClC Chloride Channels*

Jian Yin, Zhifeng Kuang, Uma Mahankali, and Thomas L. Beck*
Department of Chemistry
University of Cincinnati
Cincinnati, OH 45221-0172

*corresponding author: thomas.beck@uc.edu, 513-556-4886

## Abstract

**ClC chloride channels possess a homodimeric structure in which each monomer contains an independent chloride ion pathway. ClC channel gating is regulated by chloride ion concentration, pH, and voltage. Based on structural and physiological evidence, it has been proposed that a glutamate residue on the extracellular end of the selectivity filter acts as a fast gate. We utilize a new search algorithm which incorporates electrostatic information to explore the ion transit pathways through wild-type and mutant bacterial ClC channels. Examination of the chloride ion permeation pathways supports the proposed important role of the glutamate residue in gating. An external chloride binding site previously postulated in physiological experiments is located near a conserved basic residue adjacent to the gate. In addition, access pathways are found for proton migration to the gate, enabling pH control at hyperpolarized membrane potentials. A chloride ion in the selectivity filter is required for the pH-dependent gating mechanism.**

## INTRODUCTION

ClC chloride channels are conserved from prokaryotes to eukaryotes, and are involved in many biological functions including acidification of the stomach[1] and intracellular vesicles, excitability of skeletal muscle, and salt and water transport across epithelia.[2] Mutations in genes coding for ClC channels are associated with several diseases.[3] Recently, it has been determined that ClC channels in *E. coli* perform as anion-selective electrical shunts coupled with proton pumps which respond to extreme acid conditions.[4] Early physiological evidence from the *Torpedo* ClC-0 channel suggested a double-barreled structure which conducts with two equally-spaced open levels during burst periods separated by inactivated states.[5] The channels show evidence of multi-ion permeation.[6]

The recent determination of the crystal structure of wild-type and mutant bacterial ClC channels has given a dramatic proof of the double-barreled geometry of the channels.[7] Sequence alignment exhibits a substantial degree of conservation between bacterial and eukaryotic ClC channels; the similarity is especially strong in the selectivity filter region. Mutational studies on eukaryotic channels correlate well with the locations of key residues in the bacterial structures.[2] Thus the bacterial structures can be expected to yield insights into the functioning of eukaryotic channels.[8]

Each monomer of the *E. coli* channel contains 18 α-helices which are arranged to create an hourglass pore for chloride ion passage.[7] The distribution of α–helices is quite complex, creating an antiparallel architecture within each monomer. Intracellular and extracellular vestibules are separated by a narrow selectivity filter roughly 15 Å in length. Four highly conserved regions stabilize a chloride ion in the filter. These sequences occur at the N-termini of α-helices; the neighboring partial charges create a favorable electrostatic environment for binding of the chloride ion. The pore exhibits substantial curvature. A glutamate residue near the extracellular end of the filter blocks the pore, and Dutzler *et al.*[7] proposed that this residue gates the channel. Since the open probability and the external chloride concentration are closely coupled, it was suggested that a chloride ion replaces the glutamate gate to initiate permeation.

Following the discovery of the bacterial structure, Dutzler *et al.*[9] examined the gating mechanism in ClC channels by determining the crystal structures of two *E. coli* mutants (E148A and E148Q), where the proposed glutamate gate was altered to alanine or glutamine, respectively. In both mutants, the side chain is moved from the glutamate binding site which blocks the pore, suggesting open conformations. Three chloride ions are bound in the mutant structures, one at the previously observed filter binding site ($S_{cen}$), one near the intracellular entrance to the filter ($S_{int}$), and one near the glutamate binding site close to the extracellular end of the filter ($S_{ext}$). In conjunction with the bacterial structures, ClC-0 channels were mutated to study altered physiological behavior. The open probability increased significantly for the mutants, providing further evidence linking the glutamate to gating function. In addition, lowering the external pH for the wild-type channel increased the open probability similar to the two mutants. A gating charge close to one best fit the open probability *vs.* voltage curve; it has been proposed that the Cl ion itself is the gating charge for the ClC channels as it moves from an external binding site into a filter binding site.[6,10]

Previous work showed the important role of external Cl concentration[6,10,11] and pH[11] in the voltage-dependent gating mechanism. Increasing external chloride concentration at fixed pH shifts the open probability *vs.* voltage curves to the left (increased open probability at higher Cl concentration); depolarization leads to near unity open probabilities. Decreasing pH increases the open probability under hyperpolarization conditions, but does not shift the curve horizontally. Chen and Chen[11] developed a gating model which involves two separate voltage-dependent gating mechanisms: one which dominates for depolarization potentials and one which rationalizes the pH dependent gating at hyperpolarization. This model accurately fits the gating data.

In this paper, we examine the wild-type and mutant *E. coli* ClC channels via molecular modeling techniques to explore permeation pathways, selectivity, and the chloride concentration and pH dependences of gating (see ref. 12 for a concise summary of the current experimental work). We employ a new algorithm which exhaustively searches for favorable pathways for ion transit. The results provide a clear picture of the course of ion transit through the complex bacterial ClC channels.

## COMPUTATIONAL METHODS

We utilize two algorithms to search for ion permeation pathways. The first method was developed by Smart *et al.*[13] The algorithm requires knowledge of an initial location within the pore. Simulated annealing is employed to locate the geometric pore center (point at which the largest hard sphere can be inserted without overlap with the protein) in a horizontal plane at a given z location. The `energy' for the statistical weight is -R, where R is the distance to the edge of the nearest protein atom. The algorithm then moves up or down in the z direction and initiates a new simulated annealing step. The method generates unambiguous results for well-defined pathways without too much curvature (for example, the KcsA channel). However, due to the soft nature of the annealing potential, many pathways can be found for a more complex structure like the ClC channel, and these must be sorted based on geometric and/or electrostatic features. As an initial step, we modified the HOLE algorithm to allow for movement along pores with strong curvature. The problem of locating many candidate pores was accentuated, however, with the increased flexibility of the modified HOLE geometry-based search.

To overcome these difficulties, we have developed a new search algorithm (TransPath) which is more robust than the HOLE algorithm. Here we briefly summarize the essential features of the new algorithm; details are presented in ref. 14. We first generate a continuum dielectric model of the protein by discretization on a three-dimensional grid.[15] We numerically solve the Poisson equation either with the CHARMM[16] Poisson-Boltzmann code or with our own efficient multigrid solver.[17] Cuts in horizontal planes (parallel to the membrane plane) at several z locations are examined to locate high dielectric spots. Configurational Bias Monte Carlo[18] trajectories are initiated from each of these spots. The energy for the statistical weight includes four contributions: a geometric factor 1/R (a harsher potential than in the HOLE algorithm), a hard sphere potential between the segments of the chains, the electrostatic potential from the protein, and an external potential which nudges the biased random walk either up (extracellular) or down (intracellular). Each continuous term includes a scaling factor to yield comparable magnitudes for the various contributions during the biased random walks.

We generate swarms of Monte Carlo trajectories, and rank their importance based on the Rosenbluth weight.[18] Failed trajectories which do not make their way out of the protein are discarded. In this way, we collect an ensemble of trajectories which pass from one side of the channel to the other. Once the trajectories are generated and a representative collection is chosen based on the Rosenbluth weight, simulated annealing steps are conducted in 85-95 horizontal planes (with a plane separation of 0.5 Å) to locate the geometric pore center. The advantages of this algorithm are that no initial knowledge of the pore is required, and the annealing step is conducted *following* the generation of a statistical ensemble of viable pathways.

Generally, the paths we find collapse into one or a few prototype paths upon annealing. The method has been tested by comparison with previous results on the potassium channel[19] and with several HOLE results on the ClC mutant channels. We have found this method to be faster and more thorough than the HOLE method, and the incorporation of electrostatic information from the start aids in locating important paths for ion transit. We note that we do not include self-energy contributions in the ion potential (which would require separate Poisson solves for each ion location). These energies are always positive and will have significant values in narrow pore regions (for example, roughly 0.5 V for the gramicidin pore, ref. 20). It is shown in ref. 20 that small protein fluctuations stabilize the ions and to a large extent cancel the self-energies. For example, the fluctuations lead to nearly a 0.6 V reduction in the potential of mean force (which includes the dielectric self-energy) for a potassium ion in the gramicidin channel relative to results for a rigid channel geometry.

The located paths are single representative paths which include geometrically and electrostatically favorable features; of course the diffusional trajectories of individual ions can follow an infinite number of paths about these idealizations. The purpose of this work is to explore candidate ion transit pathways and gain insights into the relative energetics during ion passage.

## RESULTS

We first present an electrostatic map of the wild-type ClC structure from ref. 9 (fig. 1). The potential is shown in a vertical cut through the channel in a plane slightly off from center. The map shows the positive potential (blue) profile along the Cl path in the each of the monomers. The positive profile is even more distinct in the mutant E148Q (below), which is proposed to correspond to an open structure. Also, notice a large domain of negative potential (red) at the top of the channel centered in the region around the dimer interface. This negative potential arises due to several acidic residues (6 on each monomer). We will see below that this domain plays a crucial dual role: it directs chloride ions outward towards the entrance to the two anion transit pathways, and it creates a favorable electrostatic environment for penetration of protons to the glutamate gate.

Next we examine the chloride ion transit path through the mutant structures (E148A and E148Q; we label the wild-type structure E148). Illustrations of the paths appear in figs. 2a,b and the pore radii and potentials are shown in fig. 3. The radius profile illustrates the selectivity filter region (roughly 15 Å) separating the broad intracellular and extracellular vestibules. The filter region is quite narrow ($R_{min}$=1.36 Å for E148A and 1.05 Å, near the glutamine, for E148Q compared with the chloride ion radius of 1.8 Å); clearly protein fluctuations are required for the passage of anions. The same behavior is observed in the selectivity filter of the KcsA channel ($R_{min}$ = 0.78 Å from TransPath, while the potassium ion radius is 1.4 Å). The potential profile displays a very large positive region through the filter (with a maximum of 1.7 V), and broad low-level positive domains in the two vestibules arising from exposed basic residues.[21] The large binding energy in the filter is partially counteracted by a self-energy contribution (roughly 0.5-0.7V) in a continuum-level treatment. As discussed above, however, the self-energy can be largely overcome by small-amplitude protein fluctuations (ref. 20).

The large positive potential suggests, consistent with the X-ray structure, that one or more chloride ions will be bound in the filter.[8,21] Both the X-ray structure[9] and physiological experiments[6] indicate ClC channels display multi-ion conduction.[22] The chloride ion path has substantial curvature; it is directed towards the portion of the vestibules distant from the dimer interface in order to access regions of positive potential. The path confirms the initial proposal of Dutzler, *et al.*[7] and passes the three chloride binding sites (at $z = -8.73$ Å, $-2.71$ Å, and $1.01$ Å) determined from the crystal structure.[9] Both the HOLE and TransPath search algorithms locate this path. Pathway searches on the wild-type structure, on the other hand (below), exhibit a narrower restriction near the glutamate ($R_{min} = 0.71$ Å), and the electrostatic potential profile is considerably reduced in magnitude compared with the mutant profiles. Therefore, the path search results support the proposal that the two mutant structures are indeed open, and the wild-type channel is closed.

We now focus on the wild-type structure to address the mechanism of gating starting from the closed state. The experiments of Chen and Chen[11] show that the channel opens at pH = 9.6 under depolarization. This suggests that the glutamate gate can operate while charged. In Chen and Chen's two-mechanism model, this correlates with the depolarization driven rate, and carries a gating charge of approximately one. The electrostatic profile (fig. 4b) along the chloride ion path (P1 of fig. 2c) displays a positive shoulder of 0.2 V (7 kT) in the neighborhood of $z = 8$ Å. This location is very near Arg147 (and just prior to a drastic narrowing of the pore radius). We label this location $S_{bs}$ and propose that this is the external binding site suggested by Chen and Miller.[10,12] In their model the chloride ion binds to the $S_{bs}$ site with little voltage dependence, presumably because the site is located in the vestibule and is exposed to the external solution. The site attracts external chloride ions to a location near the gate.[23] The distance between the Glu148 and Arg147 side chains decreases substantially (from 6.72 Å to 3.56 Å) between E148 and E148Q. One possibility is that, during gate opening, a chloride ion at $S_{bs}$ replaces the charged glutamate in a concerted process whereby the ion fills the $S_{ext}$ location while the charged Glu148 side chain moves into closer proximity to the Arg147 side chain. Notice that the potential profile for the chloride ion path with the E148Q glutamine mutated back to a charged Glu residue still maintains a strong positive potential entering the filter (fig. 4b).

The pH-dependent hyperpolarization-driven mechanism corresponds to the second mechanism of Chen and Chen,[11] with a gating charge of approximately -0.3. This mechanism leads to increased opening rates at low pH and negative potentials. Presumably protonation of the Glu148 gate results in increased open probabilities as the neutral side chain is freed from the strong binding energy at $S_{ext}$ when charged. In order for this second mechanism to work, protons must access the Glu148 side chain. We propose that the acidic residues at the top center of the channel provide the favorable electrostatic environment required for proton propagation to the gate. Recent simulation work on proton diffusion through channels indicates the dominant role of electrostatic driving forces in proton migration through channels;[24] previous models focused on specific proton wire mechanisms. To explore the proton access paths, we converted our test charge in the TransPath search algorithm to a positive charge.

The first candidate proton path (P2, figs. 2c and 4c,d) follows the side of the extracellular vestibule nearer to the dimer interface. The computed pore is quite wide in the vestibule, and the potential in regions away from the gate is negative (-.2 V minimum near the extracellular exit, and $R_{min}$ = 1.09 Å near Arg147). While protons will diffuse along this direction to near z = 7 Å, there exists a potential barrier for penetration of the proton to Glu148 due to the Arg147 side chain. If we place chloride ions at $S_{int}$, $S_{cen}$, and $S_{bs}$, however, the potential drops considerably to large negative values (fig. 4d). If chloride ions are placed at $S_{int}$ and $S_{cen}$, an intermediate profile is obtained. Therefore, instead of the original proposal that protonation of an external site may induce chloride permeation directly,[25] the opposite process is suggested by these results; chloride binding at $S_{bs}$ alters the potential profile so a proton may access the gating region. Experimentally, the influence of external chloride ion concentration on the pH dependent gating mechanism is not entirely clear (ref. 11, fig. 7).

A second possible proton pathway (P3, figs. 2c and 4e,f) was found which coincides with the vestibular path discussed above near the extracellular entrance but deviates approaching Glu148. An acidic residue (Glu414) is located near this path which leads to a negative potential with a minimum value of -0.5 V at z =8 Å (fig. 4f). The minimum radius of this path is 0.63 Å so it is questionable whether protons associated with waters can penetrate through such a narrow region without a stronger binding potential unless there are substantial protein fluctuations. In addition, if the Glu414 residue is protonated, the potential becomes unfavorable. It is interesting to note, however, that the Glu414 residue is conserved throughout the ClC family. Both paths (P2 and P3) terminate in close proximity to Glu148.

In addition, one path was found following the dimer interface which has a minimum potential value of -0.8V at z = -3 Å but reaches positive values at more negative z locations. Therefore, this pathway likely is blocked for motion of ions of either charge. No clear access from this path to the filter region was apparent. Finally, we observed a pathway which follows the red domain in the lower portion of the channel (fig. 1); this path may possibly be involved in proton leakage through the channel (below). The path passes near three acidic residues (E113, E117, and E203).

Based on these results for the static structure, the vestibular path (P2) appears the most likely candidate for extracellular proton access to the proposed gating glutamate side chain due to its favorable electrostatic and geometric features. On the other hand, the conservation of the Glu414 residue through the ClC family suggests that it may play an important role in proton access. Protein fluctuations may allow for proton penetration through the narrower portions of the P3 path near Glu148. The (6+6) acidic residues near the channel top center are crucial for providing a negative electrostatic potential driving proton diffusion; when all of the acidic residues in this domain are protonated, the negative potential region is destroyed. Hyperpolarization provides an additional driving force for proton penetration to the gate. Mutations near P2 and/or involving the residues corresponding to Glu414 in eukaryotic channels should be able to test the relative importance of the proton access paths in gating.

Single-channel measurements have suggested the $pK_a$ of the ClC gate is roughly 5.3,[11] which implies a positive shift of about one unit. Electrostatic effects can substantially alter $pK_a$ values in a protein environment.[26] To examine the protonation state of Glu148, we performed continuum dielectric calculations on the wild-type structure.[27] The $pK_a$ shift results are presented in fig. 5. With no chloride ions in the vicinity of the glutamate, the pKa shift is large and negative (the $pK_a$ is close to zero at 0.0 V); the strong positive potential at the $S_{ext}$ binding site makes the deprotonated state stable. If two chloride ions are placed at $S_{int}$ and $S_{cen}$, the $pK_a$ value of the glutamate shifts substantially upward, to a value of around 5 (the ion at $S_{int}$ has little effect on the $pK_a$). If a third chloride ion is placed at the $S_{bs}$ location, the pKa shifts to even higher values (7.5). This suggests that protonation of Glu148 requires a chloride ion at $S_{cen}$ to create a favorable electrostatic environment for protonation. An additional chloride ion at $S_{bs}$ further enhances the probability of protonation. The important role of internal chloride concentration on the pH-dependent gating has been observed experimentally.[10,12] This appears to occur through penetration of an internal chloride ion to $S_{cen}$.

## DISCUSSION

The computational modeling yields chloride ion pathways in agreement with the initial proposal of Dutzler *et al.*[7] The radius and potential profiles for the two mutant structures support their interpretation as open structures. In addition, the results are consistent with the two-mechanism model put forward by Chen and Chen[11] for channel gating. The binding site at $S_{bs}$ attracts chloride ions which are optimally placed to enter the pore when the Glu148 side chain moves out of the way, likely in a concerted mechanism. This mechanism operates at positive voltages. At negative voltages, the gating is influenced by extracellular pH. We have located candidate proton access paths, and have shown that a chloride ion located in the filter is required to shift the $pK_a$ value of the glutamate to the appropriate range. Strategically placed acid residues at the top center of the channel both direct chloride ions outward toward the entrance to the pore and create a favorable electrostatic environment for proton access. Since the gating mechanism appears to be relatively localized, further studies should be directed at molecular dynamics simulations to investigate the glutamate motions in charged and neutral forms, and in the presence of neighboring chloride ions.

Pusch and coworkers[28] have recently questioned the local glutamate gating hypothesis of MacKinnon *et al.*[9] based on blocking studies on the intracellular side of ClC-0 channels. They found substantially different affinities of the blocker for the open and closed states, and suggested that this implies significant conformational changes (in the channel pore and filter) during gating. Part of the evidence for the larger conformational changes came from studies of the mutant Y512F which removes a hydroxyl group from the chloride binding site $S_{cen}$. While we cannot preclude conformational changes in our calculations based on fixed structures, we addressed the issue of that mutation via electrostatic modeling. First, we performed the mutation

Y445F in the E148Q mutant structure, and found virtually no change in the electrostatic profile for the chloride ion path through the filter region. This result shows the hydroxyl group is not crucial for determining the electrostatic profile through the filter. Second, the potential profile through the wild-type E148 filter region drops significantly from the E148Q profile due to the presence of the charged glutamate; the potential at $S_{cen}$ decreases from 1.409 V in E148Q to 0.545 V in E148. Therefore, it can be expected that the chloride ion occupancy at $S_{cen}$ in E148Q is substantially larger than in E148. Locating a chloride ion at $S_{cen}$ in E148Q leads to a potential of 0.099 V at $S_{int}$, relative to 0.246 V for E148. Hence, the decreased affinity for the blocker in the open state could equally well be explained by electrostatic effects due to increased occupancy of the open channel filter by chloride ions.

In this paper, we have focused on chloride ion transport mechanisms through the bacterial ClC channels, and the associated proton access to the proposed gating region. Recent work (C. Miller, personal communication) has suggested possible proton/chloride antiport behavior in the bacterial structures. Since facilitated transport requires conformational transitions in the protein, this issue has not been addressed in the present study. A possible mechanism for moving protons completely across the protein is unknown. It is interesting to note, however, the two negative potential lobes in the lower part of fig. 1, which may provide zones through which protons diffuse, perhaps associated with protein rearrangements. In preliminary computational results for a homology model of the ClC-0 channel, the negative potential at the top channel center is maintained, but the two negative potential lobes in the lower domains of the protein disappear; the lack of a negative potential domain there would prevent proton diffusion across the channel. This change in the electrostatic distribution is likely due to the replacement of the three acidic residues in the bacterial channel with basic or neutral residues in ClC-0 (E113-->K, E117-->R, E203-->V from the bacterial/ClC-0 alignment). There does exist a close correspondence between the bacterial ClC structures and observed physiological behavior of ClC-0 channels through a range of mutational studies.[2,8,9,12,29] It is likely that, while the bacterial/eukaryotic similarity may not be complete, it does capture key features of chloride ion motion through the ClC channel family.

### ACKNOWLEDGEMENTS

We gratefully acknowledge the support of the Department of Defense MURI program. We thank John Cuppoletti, Rob Coalson, Warren Dukes, and Bob Eisenberg for many helpful discussions. We especially thank Anping Liu, Director of Molecular Modeling for discussions and technical support.

**REFERENCES**
1. Stroffekova K, Kupert EY, Malinowska DH, Cuppoletti J. Identification of the pH sensor and activation by chemical modification of the ClC-2G Cl$^-$ channel. Am J Physiol 1998; 275: C1113-C1123.
2. Estevez R, Jentsch TJ. CLC chloride channels: correlating structure with function. Curr Opin Struct Biol 2002; 12: 531-539.
3. Ashcroft FM. Ion Channels and Disease. New York: Academic Press; 2000.
4. Iyer R, Iverson TM, Accardi A, Miller C. A biological role for prokaryotic ClC chloride channels. Nature 2002; 419: 715-718; see also Maduke M, Pheasant DJ, and Miller C. High-level expression, functional reconstitution, and quaternary structure of a prokaryotic ClC-type chloride channel. J Gen Physiol 1999; 114: 713-722.
5. Miller C, White MM. Dimeric structure of single chloride channels from torpedo electroplax. Proc Natl Acad Sci USA 1984; 81: 2772-2775; see also Ludewig U, Pusch M, Jentsch TJ. Two physically distinct pores in the dimeric ClC-0 chloride channel. Nature 1996; 383: 340-343.
6. Pusch M, Ludewig U, Rehfeldt A, Jentsch TJ. Gating of the voltage-dependent chloride channel ClC-0 by the permeant anion. Nature 1995; 373: 527-530.
7. Dutzler R, Campbell EB, Cadene M, Chait BT, MacKinnon R. Nature 2002; 415: 287-294.
8. Chen MF, Chen TY. Side-chain charge effects and conductance determinants in the pore of ClC-0 chloride channels. J Gen Physiol 2003; 122: 133-145.
9. Dutzler R, Campbell EB, MacKinnon R. Gating the selectivity filter in ClC chloride channels. Science 2003; 300: 108-112.
10. Chen TY, Miller C. Nonequilibrium gating and voltage dependence of the ClC-0 Cl$^-$ channel. J Gen Physiol 1996; 108: 237-250.
11. Chen MF, Chen TY. Different fast-gate regulation by external Cl$^-$ and H$^+$ of the muscle-type ClC chloride channels. J Gen Physiol 2001; 118: 23-32.
12. Chen TY. Coupling gating with ion permeation in ClC channels. Science STKE 2003; pe23.
13. Smart OS, Goodfellow JM, Wallace BA. The pore dimensions of gramicidin A. Biophys J 1993; 65: 2455-2460.
14. Kuang Z, Liu A, Yin J, Beck TL. To be submitted.
15. Dielectric constants of 4 and 80 were used for the protein and water, respectively. The calculations were performed without a surrounding membrane; some test calculations were done with a membrane which had no noticeable effect on the pore potential. The dielectric profile for the electrostatic calculations utilized the Connolly surface. Default charges were assumed. The pore radii for the hole search calculations used hard core parameters; see Turano B, Pear M, Busath D. Gramicidin channel selectivity. Molecular mechanics calculations for formamidinium, guanidinium, and acetamidinium. Biophys J 1992; 63: 152-161.

16. Brooks BR, Bruccoleri B, Olafson D, States DJ, Swaminathan S, Karplus M. CHARMM: a program for macromolecular energy minimization and dynamics calculations. J Comput Chem 1983; 4: 187-217. CHARMM v. 28 was used in our calculations.

17. Beck TL. Real-space mesh techniques in density functional theory. Rev Mod Phys 2000; 72: 1041-1080 (2000).

18. Frenkel D and Smit B. Understanding Molecular Simulation. New York: Academic Press; 1996.

19. Ranatunga KM, Shrivastava IH, Smith GR, Sansom MSP. Side-chain ionization states in a potassium channel. Biophys J 2001; 80: 1210-1219; Biggin PC, Sansom MSP. Open-state models of a potassium channel. Biophys J 2002; 83: 1867-1876.

20. Mamonov AB, Coalson RD, Nitzan A, Kurnikova MG. The role of the dielectric barrier in narrow biological channels: a novel composite approach to modeling single-channel currents. Biophys J 2003; 84: 3646-3661.

21. Lin CW, Chen TY. Probing the pore of ClC-0 by substituted cysteine accessibility method using methane thiosulfonate reagents. J Gen Physiol 2003; 122: 147-159.

22. We examined the potential profile with a chloride ion placed at $S_{cen}$. There is still a positive shoulder of magnitude 1.2 V for the E148Q structure at $S_{ext}$ and 0.7 V with the glutamine mutated to a charged Glu.

23. Lin CW, Chen TY. Cysteine modification of a putative pore residue in ClC-0. J Gen Physiol 2000; 116: 535-546.

24. Burykin A, Warshel A. What really prevents proton transport through aquaporin? Charge self-energy versus proton wire proposals. Biophys J 2003; 85: 3696-3706. See also Schirmer T, Phale PS. Brownian dynamics simulation of ion flow through porin channels. J Molec Biol 1999; 294: 1159-1167.

25. Rychkov GY, Pusch M, St J Astill D, Robers ML, Jentsch TJ, Bretag AH. Concentration and pH dependence of skeletal muscle chloride channel ClC-1. J Physiol 1996; 497: 423-435; Rychkov GY, Astill D, Bennetts B, Hughes BP, Bretag AH, Roberts ML. pH-dependent interactions of $Cd^{2+}$ and a carboxylate blocker with the rat ClC-1 chloride channel and its R304E mutant in the Sf-9 insect cell line. J Physiol 1997; 501: 355-362.

26. Honig, Nicholls A. Classical electrostatics in biology and chemistry. Science 1995; 268: 1144-1149; Berneche S, Roux B. The ionization state and the conformation of Glu-71 in the KcsA $K^+$ channel. Biophys J 2002; 82: 772-780; Fitch C, Karp DA, Lee KK, Stites WE, Lattman EE, Garcia-Moreno B. Experimental $pK_a$ values of buried residues: analysis with continuum methods and role of water penetration. Biophys J 2002; 82: 3289-3304; Sham YY, Chu ZT, Warshel A. Consistent calculations of $pK_a$'s of ionizable residues in proteins: semi-microscopic and microscopic approaches. J Phys Chem 1997; 101: 4458-4472; Bashford D, Karplus M. $pK_a$'s of ionizable groups in proteins: atomic detail from a continuum electrostatic model. Biochem 1990; 29: 10219-10225;

Demchuk E, Wade RC. Improving the continuum dielectric approach to calculating $pK_a$'s of ionizable groups in proteins. J. Phys. Chem. 1996; 100: 17373-17387; Nielsen JE, McCammon JA. Calculating $pK_a$ values in enzyme active sites. Protein Sci 2003; 12: 1894-1901; Nonner W, Eisenberg B. Ion permeation and glutamate residues linked by Poisson-Nernst-Planck theory in L-type calcium channels. Biophys J 1998; 75: 1287-1305.

27. The $pK_a$ calculations were performed with CHARMM v. 28, ref. 16. A protein dielectric constant of 10 was assumed; see, Demchuk and Wade, ref. 26. The linearized Poisson-Boltzmann equation was solved with SOR relaxation. The grid size used was 0.5 Å. A membrane dielectric constant of 2 was used. A bathing solution of 150 mM was included. The water probe radius was taken as 1.4 Å, with an ion exclusion radius 2 Å. The membrane thickness was 34 Å. Default charges were assumed. We also utilized the UHBD code, Madura JD, Briggs JM, Wade RC, Davis ME, Luty BA, Ilin A, Antosiewicz J, Gilson, MK, Bagheri B, Scott LR, McCammon JA. Electrostatics and diffusion of molecules in solution: simulations with the University of Houston Brownian Dynamics program. Comput Phys Commun 1995; 91: 57-95, with comparable parameters and results.

28. Traverso S, Elia L, Pusch M. Gating competence of constitutively open ClC-0 Mutants revealed by the interaction with a small organic inhibitor. J Gen Physiol 2003; 122: 295-306; Accardi A, Pusch M. Conformational changes in the pore of ClC-0. J Gen Physiol 2003; 122: 277-293; Moran O, Traverso S, Elia L, Pusch M. Molecular modeling of p-chlorophenoxyacetic acid binding to the ClC-0 channel. Biochem 2003; 42: 5176-5185.

29. Miller C. ClC channels: reading eukaryotic function through prokaryotic spectacles. J Gen Physiol 2003; 122: 129-131.

# Figures

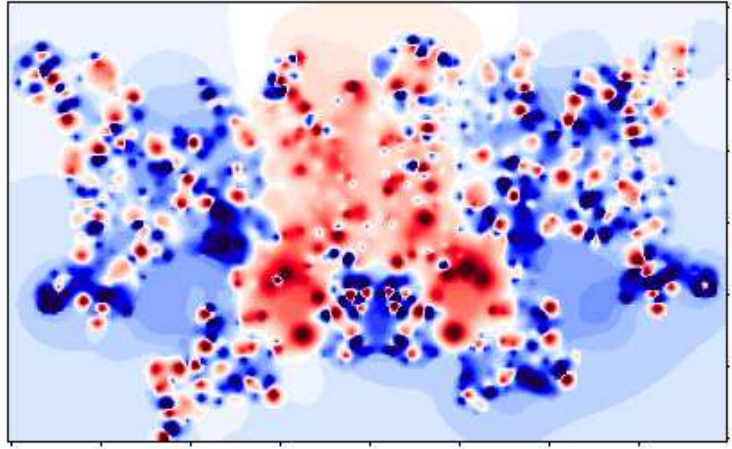

**Figure 1.** Electrostatic map of the *E. coli* wild-type structure. Blue is positive potential and red is negative potential.

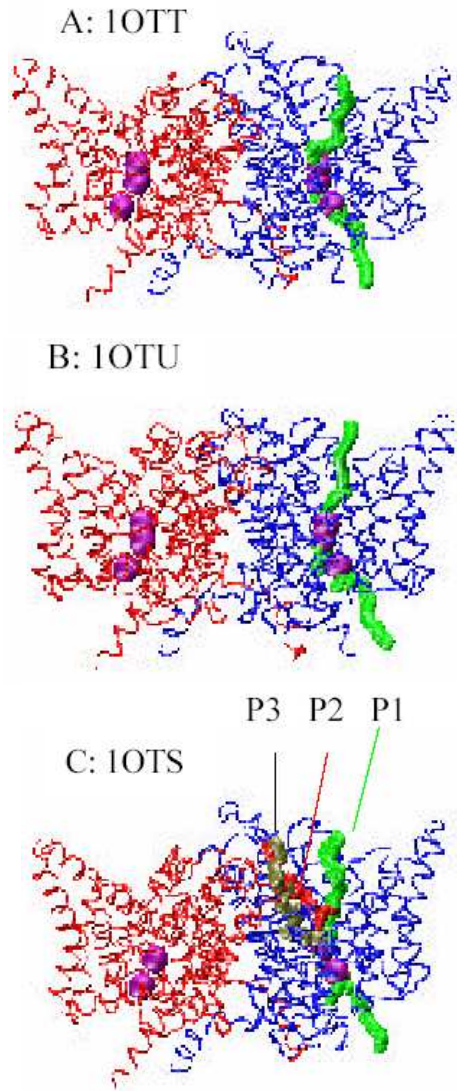

**Figure 2.** Images of the various ion transit paths. The bacterial ClC channels are shown from the side with the z direction vertical. The red and blue portions of the proteins are the two dimers composing the double-barreled channels. A) The mutant structure E148A. The purple spheres locate the three negative ion binding sites $S_{int}$, $S_{cen}$, and $S_{ext}$. The green polymer follows the chloride ion pathway. B) The mutant structure E148Q. Colors as in A). C) The wild-type structure. The three paths (P1, P2, and P3) are discussed in the text. The protein images were made with the VMD software: Humphrey W, Dalke A, Schulten K, VMD: Visual molecular dynamics. J Molec Graphics 1996; 14: 33-38.

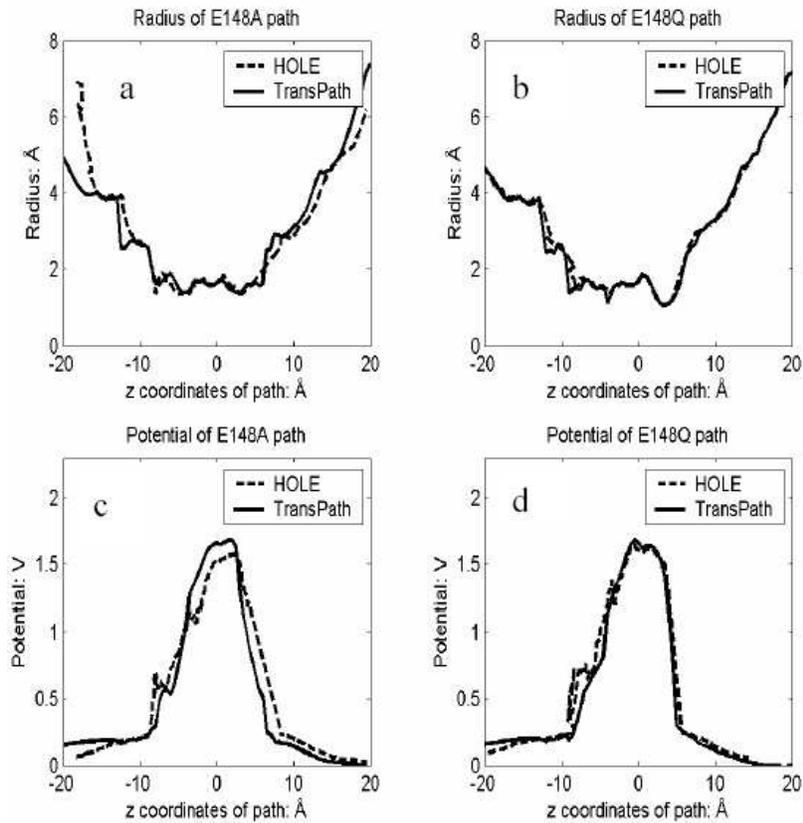

**Figure 3.** Mutant radius (a, b) and potential (c, d) profiles for chloride ion pathways. Results from HOLE and TransPath are compared. Dashed lines are the HOLE results and the solid lines are the TransPath results. E148A on left (a, c), E148Q on right (b, d).

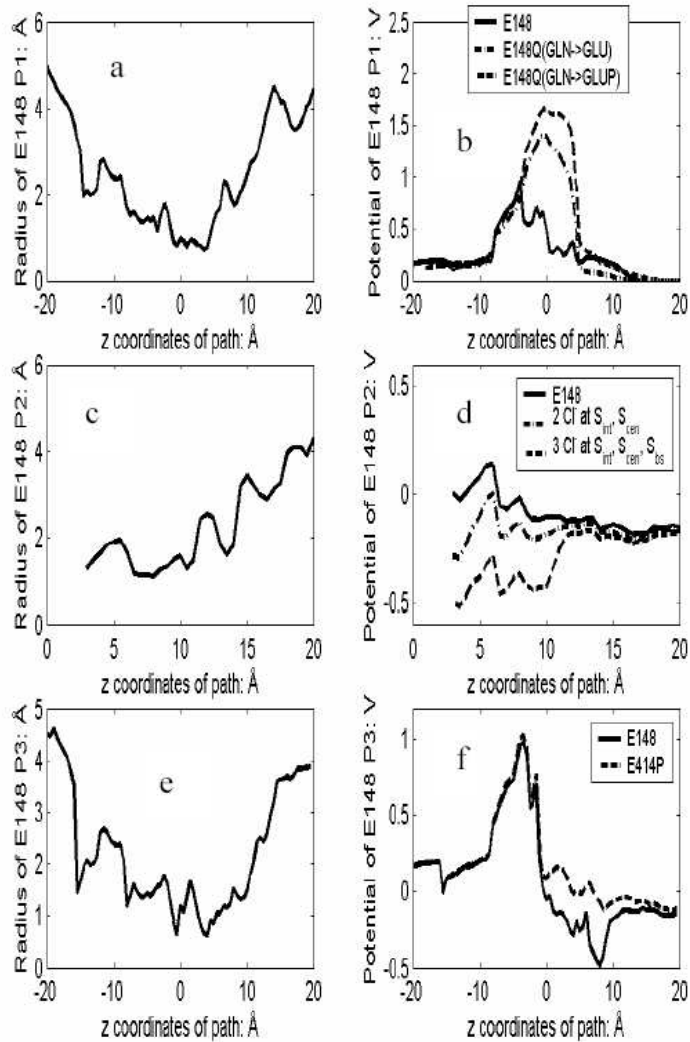

**Figure 4.** E148 pathway radii and potentials: The pore radius and potential profile for paths P1 (a, b), P2 (c, d), P3 (e, f). The top right figure (b) displays three potential profiles: the chloride path for E148 (solid), the chloride path for E148Q with the glutamine mutated to neutral Glu (dashed), the chloride path for E148Q with the glutamine mutated to charged Glu(dot/dashed). For the P2 path potential profiles (d), solid is for E148 with no chloride ions, dot/dashed is for two chloride ions, one at $S_{int}$ and one at $S_{cen}$, and dashed has an additional chloride ion at $S_{bs}$. The P3 potential profile (f) is solid, and dashed is the profile when Glu414 is neutralized.

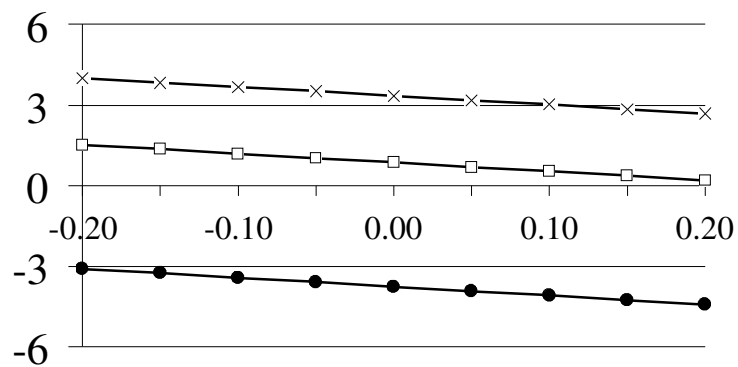

**Figure 5.** p$K_a$ shift vs. applied voltage (in volts). Filled dots are for no chlorides, open squares are for two chlorides at $S_{int}$ and $S_{cen}$. Crosses correspond to placement of an additional chloride at $S_{bs}$.